\documentclass[12pt,final,letterpaper]{annrevdistribute}

\begin{document}

\input epsf.tex
\input psfig.sty

\title{Connecting local structure to interface formation: a molecular scale
van der Waals theory of nonuniform liquids}
\author{John D. Weeks 
\affiliation{Institute for Physical Science and Technology and Department
of Chemistry and Biochemistry, University of Maryland, College Park, MD
20742 USA}}

\begin{abstract}
This article reviews a new and general theory of nonuniform fluids that
naturally incorporates molecular scale information into the classical van der
Waals theory of slowly varying interfaces. The method optimally combines two
standard approximations, molecular (mean) field theory to describe interface
formation and linear response (or Gaussian fluctuation) theory to describe
local structure. Accurate results have been found in many different
applications in nonuniform simple fluids and these ideas may have important
implications for the theory of hydrophobic interactions in water.
\end{abstract}

\date{}
\maketitle

\markboth{Weeks}{Theory of nonuniform liquids}


\thispagestyle{empty}

\newpage

\section{Introduction}

This article reviews recent progress we and our coworkers have made in
developing a new and general theory of nonuniform fluids (1-7), based on a
reexamination of the ideas that lead to the classic van der Waals (VDW)
theory (8, 9) of the liquid-vapor interface. The VDW interface theory,
developed twenty years after the VDW equation of state for the uniform
fluid, is equally far-reaching and has a great many
virtues that merit further consideration in the light of modern developments
in statistical mechanics. It is physically motivated, treating
separately the excluded volume effects associated with the short-ranged and
harshly repulsive intermolecular forces and the averaged effects of the
longer ranged attractive interactions. Both thermodynamic and structural
features are connected together naturally in an elegant and self-consistent
approach.

In principle the VDW theory can be applied to a very general problem: the
structure and thermodynamics of a fluid in the presence of a general
external field. However, the usual theory makes a crucial assumption that
the fluid density in the presence of the field is in some sense \emph{slowly
varying}. While this assumption seems appropriate for the liquid-vapor
interface in zero field where the VDW theory had its original and
spectacularly successful application (9), it fails badly when applied to the
more general and often rapidly varying fields associated with a number of
problems of current interest. By trying to understand precisely where and
why the classical theory failed, we have been able to develop a new
perspective that allows us to address these more general problems.

For example, the field can describe the interaction of fixed solutes with a
solvent liquid. The solvation free energy is related to the changes in free
energy as the interaction field is ``turned on'' from zero to full strength
(10). A significant part of the solvation free energy arises from the
required expulsion of solvent molecules from the region occupied by the
solute. This involves very strong ``excluded volume'' interactions that can
significantly perturb the density around the solutes. These considerations
also play a major role in the theory of hydrophobic interactions (4, 11-29),
and we will later review and clarify work (4) by Lum, Chandler, and Weeks
(LCW) based on this perspective. (As emphasized by Lawrence Pratt in his
review of hydrophobic interactions in this volume (30), the general theory
of hydrophobic interactions is incredibly complicated, and we will only deal
with certain important but limited aspects here.) Similar issues arise in
trying to understand effective interactions between solutes arising from
depletion forces (31).

An external field representing one or more fixed solvent particles leads to
theories for multi-particle correlations functions describing the molecular
scale structure of the bulk solvent liquid (32). Nonuniform fluids confined
by walls, slits, pores, etc., can be described using appropriate external
fields. Particular fields can enhance and alter features seen in the local
structure of the uniform fluid. Confining fields can also produce shifts and
rounding of bulk liquid-vapor and critical point phase transitions and can
change the effective dimensionality of bulk fluid correlations; e.g., from
three to two dimensional behavior in sufficiently narrow slits. In addition
the field can induce new density correlations associated with \emph{%
interface formation}, leading to \emph{wetting} or \emph{drying transitions}
and related phenomena like capillary condensation (33-42). Thus, in an
example we will consider in some detail later, a vapor-like drying region of
lower density can form near a hard wall in a Lennard-Jones (LJ)\ fluid over
a range of thermodynamic states near coexistence.
We refer to the recent review by Gelb et al
(33) for a detailed discussion of many of these possibilities.

This possible mixing of local structural components along with interfacial
components in the density response to an external field leads to the great
variety of behavior seen in nonuniform fluids. The classical VDW theory
gives a qualitatively accurate description of the smooth interfacial
components but fails for the local structure features that also must be
taken into account for even a qualitatively accurate description of problems
like those discussed above.

Of course, it has long been recognized that the VDW assumption of a slowly
varying interface is at fault. But that realization alone does not give us
much insight into how to improve things. Most modern attempts (43-60) to
address these problems have used \emph{density functional theory }(DFT).
Here one tries to express the free energy as a (generally nonlocal)
functional of the density, and the VDW theory emerges when a \emph{local
density approximation}, appropriate for a slowly varying interface, is made.
But what is that functional when the local density is rapidly varying and
even gradient corrections are inadequate? Most workers have used a
\emph{weighted density approximation} in which one averages in some way over
the rapidly varying local structure, and this approach has had great success
in some applications. However, there is no ``theory of theories'' (61) for how
to choose suitable weighting functions, and a host of different and often
highly formal schemes have been proposed. These complications stand in
strong contrast to the simplicity and physical appeal of the original VDW
theory. We will discuss DFT further in Section \ref{DFTsection}.

We show here that there is another and very simple way to derive the VDW
interface equation directly without first approximating the free energy.
This allows one to think about the ``local density'' or ``slowly
varying profile'' approximation in a new way where the theory itself
suggests what is needed to correct it. Our work can be viewed as
simply implementing these corrections to the classical VDW interface equation
and it seems appropriate to refer to it for the purposes of
this review as a \emph{molecular scale} van der Waals (MVDW) theory.

More precisely, we show that there are two key approximations in this
interpretation of the VDW theory: i) the introduction of an effective single
particle potential or ``\emph{molecular field}'' to describe the locally
averaged effects of the attractive interactions in the nonuniform fluid and
ii) the use of a \emph{hydrostatic approximation} to determine the density
response to the effective field that takes into account only the \emph{local
value} of the field. The latter approximation is accurate only for very
slowly varying fields, where it reduces to the local density approximation,
and this is what limits the utility of the VDW theory in most of the
applications discussed above. But from this perspective, it is easy to see
how the theory can be corrected by considering nonlocal effects from the
effective field. Our new MVDW theory does this using what is probably the
simplest possible method, \emph{linear response theory} (10, 62), which we
argue (5, 6) is especially accurate when used to correct the hydrostatic
approximation.

In the following, we will present this interpretation of the VDW theory,
first for the nonuniform LJ fluid. Then we will describe the corrections
and generalizations that lead to the
MVDW theory, and review results for several applications to nonuniform LJ
and hard sphere fluids, and to water. Not all the simplifications that arise
from this viewpoint were realized originally, so the present review can
serve as a simpler and more concise guide to the theory. In particular, we
have tried to clarify certain aspects of the LCW theory (4) for hydrophobic
interactions. A different view of the LCW theory that may go beyond the
MVDW picture considered originally and discussed here is given in
(63).

We first consider the LJ fluid, where the physics is particularly clear. One
reason why the LJ fluid is aptly called a simple liquid is that its local
structure at typical liquid densities can be well approximated by an even
simpler model, the \emph{hard sphere fluid}. As pointed out by Widom (64,
65), in most typical configurations of the \emph{uniform} LJ fluid the
vector sum of the longer-ranged attractive forces on a given molecule from
pairs of oppositely situated neighbors tends to \emph{cancel}. This leaves
only the excluded volume correlations induced by the harshly repulsive
molecular cores, well described by hard spheres of an appropriate size (10,
66). The hard sphere model is thus of fundamental importance in the theory
of liquids as the simplest model that can give a realistic description of
structural correlations arising from excluded volume effects.

As emphasized by VDW himself (9), this cancellation argument
must fail for \emph{nonuniform} fluids,
and there will exist ``unbalanced'' attractive forces, whose effect on the
structure can also be quite significant, leading in some cases to interface
formation. In general there is a \emph{competition} between
these two sources of
structural correlations that a proper theory of nonuniform fluids should
account for (1-7). This physical picture will help us understand what is
missing in the classical VDW theory and how it might be improved. The
insights gained from a careful study of this simple system may suggest
physically well-grounded approximations that could be applied more
generally. We believe this is the case for the MVDW theory.

We start by defining the interactions in the nonuniform LJ fluid. Particles
interact with a known external field $\phi (\mathbf{r})$, which we will
initially assume is nonzero only in a local region, e.g., the potential
arising when a LJ or hard core solute is fixed at the origin. We describe
the system using a \emph{grand ensemble}
with fixed chemical potential $\mu ^{B}$,
which determines $\rho ^{B}$, the uniform bulk fluid density far from the
perturbation where $\phi (\mathbf{r})=0.$ The LJ pair potential $w(r)\equiv
u_{0}(r)+u_{1}(r)$ is separated into rapidly and slowly varying parts
associated with the intermolecular \emph{forces} (67-69) so that all the
harshly repulsive forces arise from $u_{0}$ and all the attractive forces
from $u_{1}.$ Thus $u_{0}(r)=w(r)+\epsilon $ for $r\leq r_{0},$ where $%
r_{0}\equiv 2^{1/6}\sigma $ is the distance to the potential minimum where
the LJ force changes sign, and is zero otherwise, while $u_{1}(r)=-\epsilon $
for $r\leq r_{0}$ and equals $w(r)$ otherwise. Here $\sigma $ and $\epsilon $
are the usual length and energy parameters in the LJ potential. With this
separation $u_{1}(r)$ is relatively slowly varying and smooth, with a
continuous derivative even at $r_{0}$. We will make use of these features in
the theory described below.

\section{Molecular field approximation}

\subsection{Structure of the nonuniform reference fluid}

We now begin our derivation of the VDW interface equation.
The fundamental approximation in this interpretation of
the VDW theory is the introduction of an effective
single particle potential or ``molecular field'' that describes the locally
averaged effects of the attractive interactions (8) in the nonuniform LJ
fluid. Since the attractive interactions are relatively slowly varying, such
an averaged treatment seems physically reasonable. The structure of the
resulting ``reference'' or ``mimic'' system, where the attractive intermolecular
interactions have been replaced by the effective single particle potential, is
supposed to accurately approximate that of the original system.
Thus the structure of the nonuniform LJ fluid is assumed to be given by
that of the simpler \emph{nonuniform reference fluid} (1-3),
with only repulsive intermolecular pair
interactions $u_{0}(r)$ (equal to the LJ repulsions) and a chemical
potential $\mu _{0}^{B}$ corresponding to the same bulk density $\rho ^{B}$
but in a different renormalized or \emph{effective reference field} (ERF) $%
\phi _{R}(\mathbf{r}).$ Because $u_{0}$ is harshly repulsive, many
properties of the reference fluid can be accurately approximated by those of
a fluid of hard spheres with a diameter chosen by the usual ``blip
function'' expansion (10, 66), as described in detail in (5, 6). While this
approximation is not essential, it is numerically very convenient, and for
most purposes we will treat the reference system as a hard sphere fluid in
the presence of the ERF.

Before we discuss the specific molecular field equation \ref{mfeqn} usually
used to determine the ERF, let us consider some general physical consequences
arising from the use of any ERF to describe the effects of attractive
interactions that will be important in what follows. Since the goal is to
produce structure in the reference fluid approximating that of the full fluid
to the extent possible, it is natural to choose $\phi _{R}(\mathbf{r})$ in
principle so that the local (singlet) densities (70) at every point $\mathbf{%
r}$ in the two fluids are equal: 
\begin{equation}
\rho _{0}(\mathbf{r};[\phi _{R}],\mu _{0}^{B})=\rho (\mathbf{r};[\phi ],\mu
^{B}).  \label{singletden}
\end{equation}
Of course this density is not known in advance, so in practice we will make
approximate choices for $\phi _{R}$ motivated by molecular field ideas. Here
the subscript $0$ denotes the reference fluid, the absence of a subscript
the LJ fluid, and the notation $[\phi ]$ indicates that the correlation
functions are functionals of the external field. Unless we want to emphasize
this point, we will suppress this functional dependence, e.g., writing
Equation \ref{singletden} as $\rho _{0}(\mathbf{r)=}\,\rho (\mathbf{r).}$

But the reference fluid can also describe certain features of higher order
correlation functions. We expect that when $\phi _{R}$ is chosen so that
Equation \ref{singletden} holds, this will produce similar local
environments for the (identical) repulsive cores in the two fluids, which
at high density will mainly determine higher order
density correlations through excluded
volume effects. Thus, when Equation \ref{singletden} is satisfied, it seems
plausible that pair correlations are also approximately equal (1, 2), so that 
\begin{equation}
\rho ^{(2)}(\mathbf{r}_{1},\mathbf{r}_{2}) \simeq \rho _{0}^{(2)}(\mathbf{r%
}_{1},\mathbf{r}_{2})\,.  \label{pairden}
\end{equation}
In the dense uniform LJ fluid it is well known that correlations are
dominated by the repulsive forces, and this near equality lies behind the
success of perturbation theories of liquids (10, 68, 69). (However, this structural
assumption is \emph{rigorously} true only in the artificial ``Kac limit''
where the attractive interactions are infinitely weak and long-ranged (64,
71).) The most general use of this idea asserts that \emph{all} structural
effects of attractive forces in the nonuniform LJ fluid can be approximately
described in terms of the structure of the reference fluid in the appropriately
chosen ERF.

There are many advantages in using the simpler reference system to define
structure. In particular, the uniform reference fluid is well defined for
all densities from dilute vapor to dense fluid and there are no conceptual
problems that arise from densities in the two phase region, as would be the
case in the traditional
theories requiring such densities in the original LJ fluid. See,
e.g., Section \ref{classicalvdw} below. Moreover, as we will discuss in
detail later, there exist simple and accurate theories for the reference
fluid structure based on linear response theory.

However there are inherent limitations in such an approximation, most
notably in describing long wavelength correlations such as those found near
the critical point. Some other shortcomings of the molecular field
approximation when applied to realistic fluid models are easily understood.
For example, the effects of long wavelength capillary wave fluctuations (72)
of the free liquid-vapor interface clearly cannot be described using
such reference system correlation functions
however $\phi _{R}$ is chosen.

\subsection{Simple molecular field equation}

\label{simplemf}We now turn to the choice of $\phi _{R}.$
In our interpretation, the classical VDW theory uses the
simple \emph{molecular field }(MF) approximation for
the ERF, given in Equation \ref{mfeqn} below.
This is just a transcription of the usual molecular field equation
for the Ising model to a continuum fluid with attractive interactions $%
u_{1}(r)$ and can be arrived at in a number of different ways (8, 44).
Particularly relevant for our purposes here is the derivation discussed in
detail in (1-3, 7) that starts from the balance of forces in the reference
and LJ fluids as described by the exact Yvon-Born-Green (YBG) hierarchy (10)
and uses Equations \ref{singletden} and \ref{pairden} to arrive at the MF
equation by an approximate integration. This can be looked on as a modern
version of the closely related calculation VDW originally carried out (9).
The final result is well known:
\begin{equation}
\phi _{R}(\mathbf{r}_{1})=\phi (\mathbf{r}_{1})+\int d\mathbf{r}_{2}\,\rho
_{0}(\mathbf{r}_{2};\mathbf{[}\phi _{R}],\mu _{0}^{B})\,u_{1}(r_{12})+2a\rho
^{B},  \label{mfeqn}
\end{equation}
where 
\begin{equation}
a\equiv -\frac{1}{2}\int d\mathbf{r}_{2}\,u_{1}(r_{12})  \label{aint}
\end{equation}
corresponds to the attractive interaction parameter $a$ in the uniform fluid
VDW equation, as discussed below. The last term in Equation \ref{mfeqn}
represents a constant of integration in the derivation in (1-3, 7) and is
chosen so that $\phi _{R}$ vanishes far from a localized perturbation where
the density becomes equal to $\rho ^{B}.$

Because of the integration over the slowly varying attractive component of the
intermolecular potential $u_{1}(r_{12}),$ the second term
on the right in Equation \ref{mfeqn} is smooth and relatively slowly varying
even when $\rho _{0}$ itself has discontinuities and oscillations, as could
arise from a $\phi $ with a hard core.
Thus the ERF $\phi _{R}(\mathbf{r}_{1})$ is quite
generally comprised of the original field $\phi (\mathbf{r}_{1})$ and
an additional \emph{slowly varying} term
\begin{equation}
\phi _{s}(\mathbf{r}_{1})\equiv \int d\mathbf{r}_{2}\,\rho _{0}(\mathbf{r}%
_{2})\,u_{1}(r_{12})+2a\rho ^{B}  \label{phis}
\end{equation}
that takes account of spatial variations of the attractive interactions,
i.e., the \emph{unbalanced attractive forces }in the nonuniform LJ fluid.

Some thermodynamic implications of the MF approximation can be seen when we
consider the MF equation \ref{mfeqn} in the limit of a \emph{constant} field.
In the grand ensemble, this is equivalent to a shift of the chemical
potential, producing a shift in the uniform density, as discussed in more
detail in the next Section. Thus we can use Equation \ref{mfeqn} to relate
chemical potentials in the reference and LJ systems.
Let $\mu (\rho )$ and $\mu _{0}(\rho )$
denote the chemical potential as a function of density in the uniform LJ
system and reference system respectively. (These also depend on the
temperature, but we are usually interested in density variations along
particular isotherms, so we will not indicate the temperature dependence
explicitly.) Then $\phi ^{\rho }=$ $\mu ^{B}-\mu (\rho )$ is the exact value
of that constant field such that the density in the LJ fluid changes from
$\rho ^{B}$ to $\rho ,$ and $\phi _{R}^{\rho }=\mu _{0}^{B}-\mu _{0}(\rho )$
is the analogous field in the reference fluid producing the same density
change. Using Equation \ref{singletden}, the MF equation \ref{mfeqn} gives
the MF \emph{approximation }relating $\phi ^{\rho }$ and $\phi _{R}^{\rho },$
or equivalently, the MF approximation relating $\mu(\rho )$
and $\mu _{0}(\rho )$. This can be written in the familiar VDW form (64): 
\begin{equation}
\mu (\rho )=\mu _{0}(\rho )-2a\rho .  \label{mumf}
\end{equation}

In the limit of a uniform system, Equation \ref{mfeqn} describes all effects
of attractive interactions in terms of the \emph{constant} parameter $a$.
Indeed the theory then reduces to the generalized uniform fluid VDW theory
of Longuet-Higgins and Widom (64, 65), where one combines an accurate
description of the uniform (hard sphere) reference system with the simple
treatment of the attractive interactions in terms of a constant background
potential $a.$ In the MF approximation $\mu (\rho )$ is \emph{defined} by
the right side of Equation \ref{mumf} and has meaning even for values of $%
\rho $ in the two phase region.

In general, to determine $\phi _{R}$ a self-consistent solution of Equation 
\ref{mfeqn} is required, since $\phi _{R}$ appears explicitly on the left
side and implicitly on the right side through $\rho _{0}(\mathbf{r};\mathbf{[%
}\phi _{R}],\mu _{0}^{B})$. In principle, since a unique density $\rho _{0}(%
\mathbf{r};\mathbf{[}\phi _{R}],\mu _{0}^{B})$ is associated with a given
external field $\phi _{R}$ through the partition function, Equation \ref
{mfeqn} is self-contained and hence self-consistent values for both $\phi _{R}$
and $\rho _{0}$ can be found, by iteration, for example. Such solutions were
found in (1, 2), using computer simulations to accurately determine the
associated density $\rho _{0}(\mathbf{r};\mathbf{[}\phi _{R}],\mu _{0}^{B})$
for a variety of external fields.

In practice, one must make additional approximations beyond the molecular
field assumption to obtain $\rho _{0}(\mathbf{r};\mathbf{[}\phi _{R}],\mu
_{0}^{B})$ in an accurate and computationally practical way. It is here that
the main limitation of the classical VDW theory arises. 
In our derivation the classical VDW theory results when a
\emph{second }approximation, appropriate for a slowly varying density
field (8), is used to determine the density
$\rho _{0}(\mathbf{r};\mathbf{[}\phi _{R}],\mu _{0}^{B})$ induced by
$\phi _{R}$. This approximation takes account only of the local value of
the field through a shifted chemical potential and can be used for a general
$\phi _{R}(\mathbf{r})$. When $\phi _{R}(\mathbf{r})$ is very slowly varying
it is exact. But in more general cases it gives inaccurate results, spoiling
the predictions of the VDW theory. To see how this comes about, we first
define this local \emph{hydrostatic density response} for a general field,
and then show how its use
transforms Equation \ref{mfeqn} into the VDW interface equation as it is
usually presented.

\section{Hydrostatic density}

\subsection{Local response to general field}

Consider first a given general external field $\phi _{R}$.
Since the chemical potential acts like a constant field in the grand partition
function, the associated density
$\rho _{0}(\mathbf{r};[\phi _{R}],\mu _{0}^{B})\equiv \rho _{0}(%
\mathbf{r)}$ is a functional of $\phi _{R}$ and a function of $\mu _{0}^{B}$
and depends only on the \emph{difference} between these quantities.
Thus for any fixed position $\mathbf{r}_{1}$ we can define a
\emph{shifted chemical potential}
$\mu _{0}^{\mathbf{r}_{1}}\equiv \mu _{0}^{B}-\phi _{R}(\mathbf{r}_{1})$
and \emph{shifted field} $\phi _{R}^{\mathbf{r}%
_{1}}(\mathbf{r})\equiv \phi _{R}(\mathbf{r})-\phi _{R}(\mathbf{r}_{1})$,
whose parametric dependence on $\mathbf{r}_{1}$ is denoted by a superscript,
and we have for all $\mathbf{r}$ the \emph{exact} relation $\rho _{0}(%
\mathbf{r};\mathbf{[}\phi _{R}],\mu _{0}^{B})=\rho _{0}(\mathbf{r};\mathbf{[}
\phi _{R}^{\mathbf{r}_{1}}],\mu _{0}^{\mathbf{r}_{1}}).$

By construction the shifted field $\phi _{R}^{\mathbf{r}_{1}}(\mathbf{r})$
vanishes at $\mathbf{r}=$ $\mathbf{r}_{1}$. If $\phi _{R}$ is \emph{very
slowly varying}, then it remains very small for $\mathbf{r}$ near $\mathbf{r}
_{1}$. In that case, to determine the density at $\mathbf{r}_{1}$ we can
approximate $\rho _{0}(\mathbf{r}_{1};\mathbf{[}\phi _{R}^{\mathbf{r}%
_{1}}],\mu _{0}^{\mathbf{r}_{1}})$ by $\rho _{0}(\mathbf{r}_{1}\mathbf{;}%
\mathbf{[}0],\mu _{0}^{\mathbf{r}_{1}})\equiv \rho _{0}^{\mathbf{r}_{1}}$,
the density of the \emph{uniform} fluid (in zero field) at the shifted
chemical potential $\mu _{0}^{\mathbf{r}_{1}}.$ This defines $\rho _{0}^{%
\mathbf{r}_{1}}$, the \emph{hydrostatic density response} to the field $\phi
_{R}.$ Note that $\rho _{0}^{\mathbf{r}_{1}}$ depends only on the \emph{local%
} value of the field $\phi _{R}$ at $\mathbf{r}_{1}$ through its dependence
on $\mu _{0}^{\mathbf{r} _{1}}.$ When the field varies slowly enough the 
\emph{hydrostatic approximation}, where $\rho _{0}(\mathbf{r}_{1})$
at each $\mathbf{r}_{1}$ is replaced by the corresponding uniform fluid
density $\rho _{0}^{\mathbf{r}_{1}}$, is very accurate (5).

One can equivalently define the hydrostatic density $\rho _{0}^{\mathbf{r}
_{1}}$ using $\mu _{0}(\rho ),$ the chemical potential of the uniform
reference fluid as a function of density; it is defined by
\begin{equation}
\mu _{0}(\rho _{0}^{\mathbf{r}_{1}})=\mu _{0}^{B}-\phi _{R}(\mathbf{r}_{1}).
\label{hydromudef}
\end{equation}
This equation plays a central role in all that follows. The hydrostatic
density is easy to determine for a general $\phi _{R}.$ In particular if
$\phi _{R}$ has a hard core with $\phi _{R}=\infty $
in a certain region of space, then the
hydrostatic density $\rho _{0}^{\mathbf{r}_{1}}$ correctly vanishes in that
same region, corresponding to the vanishing density of the uniform fluid at
the chemical potential $\mu _{0}^{\mathbf{r}_{1}}=-\,\infty $. However,
because of the strictly local response to the field, any nonlocal excluded
volume correlations induced by the hard core potential outside the hard core
region are not properly described by the hydrostatic approximation. Still,
it remains well defined even in this limit whereas approximations based on using
properties of a uniform fluid evaluated at the \emph{local density} (which can
easily exceed close packing) break down completely (43, 44).

\subsection{Local response to ERF}

\label{responsetoERF}To obtain the VDW and MVDW theories we now determine
the local hydrostatic response to the ERF in Equation \ref{mfeqn}.
This arises after Equation \ref{mfeqn} for $\phi _{R}(\mathbf{r}%
_{1})$ is substituted into Equation \ref{hydromudef}: 
\begin{equation}
\mu _{0}(\rho _{0}^{\mathbf{r}_{1}})=\mu _{0}^{B}-\phi (\mathbf{r}_{1})-\int
d\mathbf{r}_{2}\,\rho _{0}(\mathbf{r}_{2})\,u_{1}(r_{12})-2a\rho ^{B}.
\label{hydroerf}
\end{equation}
As written, Equation \ref{hydroerf} just defines the hydrostatic density $%
\rho _{0}^{\mathbf{r}_{1}}$ in terms of the local value of the ERF at
$\mathbf{r}_{1}$, which itself involves an integral over the full density
$\rho _{0}(\mathbf{r}_{2};[\phi _{R}],\mu _{0}^{B})$
at all other points $\mathbf{r}_{2}.$ To explicitly determine
$\rho _{0}^{\mathbf{r}_{1}}$ we need to specify $\Delta \rho _{0}(%
\mathbf{r}_{2})\equiv \rho _{0}(\mathbf{r}_{2})-\rho _{0}^{\mathbf{r}_{2}},$
as is clear when Equation \ref{hydroerf} is exactly rewritten in the form: 
\begin{eqnarray}
\mu _{0}(\rho _{0}^{\mathbf{r}_{1}})-2a\rho _{0}^{\mathbf{r}_{1}} &=&\mu
_{0}^{B}-2a\rho ^{B}-\phi (\mathbf{r}_{1})-\int d\mathbf{r}_{2}\,[\rho _{0}^{%
\mathbf{r}_{2}}\,-\rho _{0}^{\mathbf{r}_{1}}]u_{1}(r_{12})  \nonumber \\
&&-\int d\mathbf{r}_{2}\Delta \rho _{0}(\mathbf{r}_{2})\,u_{1}(r_{12}).
\label{hydroerf1}
\end{eqnarray}
As we now show, different approximations for $\Delta \rho _{0}(\mathbf{r}_{2})$
immediately give the classical VDW and the MVDW interface equations.

\section{Hydrostatic approximation and the classical VDW theory}

\label{classicalvdw}The classical VDW interface equation arises when one
assumes that the full
density response to the ERF $\rho_{0}(\mathbf{r}_{2})$
for all $\mathbf{r}_{2}$ is accurately approximated by 
$\rho _{0}^{\mathbf{r}_{2}}$, the local hydrostatic response. This
hydrostatic approximation is consistent when the ERF and the induced density
are slowly enough varying. Thus in this derivation of the classical VDW
theory the last term in Equation \ref{hydroerf1} is ignored and Equation
\ref{hydroerf} or \ref{hydroerf1} is approximated  by an integral equation
involving only the hydrostatic density:
\begin{equation}
\mu (\rho _{0}^{\mathbf{r}_{1}})=\mu ^{B}-\phi (\mathbf{r}_{1})-\int d%
\mathbf{r}_{2}\,[\rho _{0}^{\mathbf{r}_{2}}\,-\rho _{0}^{\mathbf{r}%
_{1}}]u_{1}(r_{12}),  \label{vdwref}
\end{equation}
along with the assumption that
$\rho _{0}(\mathbf{r})=\rho _{0}^{\mathbf{r}}$ for all $\mathbf{r}$.
Here we have also used Equation \ref{mumf} to replace $\mu _{0}(\rho )$ by $%
\mu (\rho ),$ with the understanding that the latter is really defined by
the right side of Equation \ref{mumf}. Consistent with the assumption of a
slowly varying profile and the fact that $u_{1}$ is reasonably short-ranged, 
$\rho _{0}^{\mathbf{r}_{2}}$ in the last term on the right is often expanded
to second order in a Taylor series about $\rho _{0}^{\mathbf{r}_{1}}$,
yielding in the simple case of a liquid-vapor interface with planar symmetry
and $\phi =0$ a differential equation for the interface profile $\rho
_{0}^{z}$: 
\begin{equation}
\mu (\rho _{0}^{z})=\mu ^{B}+m\,d^{2}\rho _{0}^{z}/dz^{2}  \label{vdwrefode}
\end{equation}
where 
\begin{equation}
m\equiv -\frac{1}{6}\int d\mathbf{r\,}r^{2}u_{1}(r).  \label{vdwm}
\end{equation}

While there is a strictly local response to the value
of the ERF at $\mathbf{r}_{1}$ in
Equation \ref{vdwref} yielding $\rho _{0}^{\mathbf{r}_{1}},$
the ERF itself at $\mathbf{r}_{1}$ depends nonlocally on $\rho _{0}^{\mathbf{%
r}_{2}}$ through the integration over the attractive interactions. In the
classical VDW theory this provides the \emph{only} source of nonlocality.

Equations \ref{vdwref} and \ref{vdwrefode} are completely equivalent to the
VDW theory for the density profile as it is usually presented (8). In our
derivation the theory describes hydrostatic densities in the reference
system, and $\mu (\rho )$ is also defined in terms of reference system
quantities given on the right side of Equation \ref{mumf}. This provides a
simple and consistent interpretation that avoids all problems associated with
densities in the two phase region of the LJ fluid.

However, in view of Equation \ref {singletden}, one can replace
$\rho _{0}^{\mathbf{r}_{1}}$ by $\rho ^{\mathbf{r}_{1}}$ in Equations
\ref{vdwref} and \ref{vdwrefode} and formally eliminate all explicit mention
of the reference system. This is the way the standard theory is usually
presented. This formal rewriting of Equations \ref{vdwref} and
\ref{vdwrefode} seems to require only properties of the original LJ system
and could be useful when applying the theory to other fluids where the
appropriate reference system is not so apparent. Work for water (4) along
these lines will be reviewed later. However, this obscures some of the
physical underpinnings of the theory in terms of the basic MF approximation
and it is not clear from the form of these equations how they should be
corrected in cases where the density profile is more rapidly varying.

Of course, one might hope that there
could exist a more general formulation of the theory where one does not need
the MF approximation at all. Despite some formal results from density
functional theory, which we will discuss later, we believe in practice this
is not likely to be the case. Long-standing conceptual problems (8) arise in
interpreting what is meant by $\mu (\rho ^{\mathbf{r}_{1}})$ and $\rho ^{%
\mathbf{r}_{1}}$ itself in Equations \ref{vdwref} and \ref{vdwrefode} for
density values in the two phase region of the uniform LJ fluid unless they are
defined by using a MF approximation either explicitly as in the right hand side
of Equation \ref{mumf}, or implicitly by assuming some kind of analytic
continuation of values from the stable phases. See Section \ref{improvevdw}
below for further discussion.

Thus to improve the classical VDW theory, we return to Equations \ref
{hydroerf} and \ref{hydroerf1}. Here the essential physics involving the
interplay between the MF approximation for the ERF and the hydrostatic
density response is clear. If we can understand in detail how to improve the
VDW theory for the LJ system, we may gain insights that could apply more
generally.

\section{Correcting the hydrostatic approximation}

\subsection{Optimized linear response and the HLR equation}

As emphasized in Section \ref{responsetoERF}, the VDW theory determines the
local density response $\rho _{0}^{\mathbf{r}_{1}}$ to the ERF $\phi _{R}(%
\mathbf{r}_{1})$ in Equations \ref{hydroerf} and \ref{hydroerf1}. Problems
arise in the classical VDW theory from the second approximation of
ignoring the density difference
$\Delta \rho _{0}(\mathbf{r}_{2})\equiv \rho _{0}(\mathbf{r}_{2})-\rho _{0}^{%
\mathbf{r}_{2}}$, produced by
definition from nonzero values of the shifted field $\phi _{R}^{\mathbf{r}%
_{2}}(\mathbf{r})$ away from $\mathbf{r}_{2}$.

But one can calculate $\Delta \rho _{0}$ in a very simple way by using \emph{%
linear response theory}, which exactly relates \emph{small} changes in the
density and field. This approach is clearly correct when $\Delta \rho _{0}$
and $\phi _{R}^{\mathbf{r}_{2}}$ are uniformly small, and
in this sense is analogous to a gradient correction to the local
density approximation in DFT. However, in contrast to the gradient
correction, we will see that there are good physical reasons to believe that
our linear response treatment could remain accurate
even for large perturbations (3, 5, 6). This will allow us to develop a new
and generally accurate theory for the nonuniform reference fluid in an
arbitrary external field and also will help us correct one of the major
shortcomings of the classical VDW theory for fluids with attractive
interactions.

We start from the general linear response equation (10) for a nonuniform
system in a \emph{general} field $\phi _{R}$, with chemical potential
$\mu _{0}^{B}$, inverse temperature $\beta =(k_{B}T)^{-1}$ and associated
density $\rho _{0}(\mathbf{r})$: 
\begin{equation}
-\beta \delta \phi _{R}(\mathbf{r}_{1})=\int \!d\mathbf{r}_{2}\,\chi
_{0}^{-1}(\mathbf{r}_{1},\mathbf{r}_{2};\mathbf{[}\rho _{0}])\delta \rho
_{0}(\mathbf{r}_{2}),  \label{linresponse}
\end{equation}
which relates small perturbations in the density and potential through the
(inverse) linear response function 
\begin{equation}
\chi _{0}^{-1}(\mathbf{r}_{1},\mathbf{r}_{2};\mathbf{[}\rho _{0}])\equiv
\delta (\mathbf{r}_{1}\!-\!\mathbf{r}_{2})/\rho _{0}(\mathbf{r}%
_{1})\!-\!c_{0}(\mathbf{r}_{1},\mathbf{r}_{2};\mathbf{[}\rho _{0}]).
\label{chiinverse}
\end{equation}
Here $c_{0}$ is the direct correlation function of the system with density 
$\rho _{0}(\mathbf{r}).$ In most cases we will consider perturbations about
a uniform system, so $\!c_{0}$ will take the simple form $%
\!c_{0}(r_{12;}\rho ).$ Since we want to focus on effects of the perturbing
field, we have used the inverse form of linear response theory, where the
field appears explicitly only on the left hand side of Equation \ref
{linresponse}, evaluated at $\mathbf{r}_{1}.$

Could Equation \ref{linresponse} also be used to determine the finite
density response $\Delta \rho _{0}$ to a \emph{large} field perturbation
$\Delta \phi _{R}$? Such a linear relation between
a (possibly infinite) external
field perturbation on the left hand side and the finite induced density
change on the right must certainly fail for values of $\mathbf{r}_{1}$ where
the field is very large. Conversely, Equation \ref{linresponse} should be
most accurate for those values of $\mathbf{r}_{1}$ where the field is small
--- in particular where the field \emph{vanishes} --- and then through the
integration over all $\mathbf{r}_{2}$ it relates density changes in the
region where the field vanishes to density changes in the regions where the
field is nonzero (3, 5).

This optimal condition for the validity of linear response theory holds true
\emph{automatically} when we use Equation \ref{linresponse} to
determine the change
at each $\mathbf{r}_{1}$ from the uniform fluid with density $\rho _{0}^{%
\mathbf{r}_{1}}$ induced by the shifted field. Thus we set $\chi
_{0}^{-1}=\chi _{0}^{-1}(r_{12};\rho _{0}^{\mathbf{r}_{1}})$ in Equation \ref
{linresponse} and take $\delta \phi _{R}=\phi _{R}^{\mathbf{r}_{1}}$ and $%
\delta \rho _{0}(\mathbf{r}_{2})=\rho _{0}(\mathbf{r}_{2})-\rho _{0}^{%
\mathbf{r}_{1}}$. Since $\phi _{R}^{\mathbf{r}_{1}}(\mathbf{r})$ is zero at $%
\mathbf{r}_{1}$ by construction, the left side of Equation \ref{linresponse}
vanishes, and we have 
\begin{equation}
0=\int \!d\mathbf{r}_{2}\,\chi _{0}^{-1}(r_{12};\rho _{0}^{\mathbf{r}%
_{1}})[\rho _{0}(\mathbf{r}_{2})-\rho _{0}^{\mathbf{r}_{1}}],
\label{chilinhydro}
\end{equation}
which can be rewritten using Equation \ref{chiinverse} as 
\begin{equation}
\rho _{0}(\mathbf{r}_{1})=\rho _{0}^{\mathbf{r}_{1}}+\rho _{0}^{\mathbf{r}%
_{1}}\int \!d\mathbf{r}_{2\,}c_{0}(r_{12};\rho _{0}^{\mathbf{r}_{1}})[\rho
_{0}(\mathbf{r}_{2})-\rho _{0}^{\mathbf{r}_{1}}].  \label{HLR}
\end{equation}

This is our final result, which we refer to as the \emph{hydrostatic linear
response} (HLR) equation. A more formal derivation involving a coupling
parameter integration, and yielding another related equation for $\rho _{0}(%
\mathbf{r}_{1})$ is given in (5). Note that the field appears only
implicitly through its local effect on $\rho _{0}^{\mathbf{r}_{1}}.$ The HLR
equation is useful by itself as a way to determine the density change
induced by a known external field $\phi _{R}(\mathbf{r}).$ It builds on
properties of the \emph{uniform} reference system, requiring in particular $%
c_{0}(r_{12};\rho )\ $and $\mu _{0}(\rho )$. Quantitatively accurate and
computationally convenient approximations for these functions are known from
the GMSA theory of Waisman (73, 74); in many cases the simpler Percus-Yevick
(PY) approximation (10, 62) discussed below gives sufficient accuracy. Given
these functions and a known external field $\phi _{R}(\mathbf{r})$, Equation 
\ref{HLR} is a \emph{linear} integral equation that can be solved
self-consistently to determine the induced density $\rho _{0}(\mathbf{r}%
_{1}) $ for all $\mathbf{r}_{1}.$ We will first discuss some general
properties of the HLR equation, and examine its accuracy in several test
cases where the external field $\phi _{R}(\mathbf{r}_{1})$ is known.
Then in Section \ref{twostep} we will discuss its
use in correcting the classical VDW
theory where $\phi _{R}(\mathbf{r}_{1})$ is determined self-consistently.

\subsection{Properties of the HLR equation for fixed $\phi _{R}$}

Equation \ref{HLR} relates the density $\rho _{0}(\mathbf{r}_{1})$ at
each $\mathbf{r}_{1}$ to an integral involving the density $\rho _{0}(%
\mathbf{r}_{2})$ at all other points and a locally optimal uniform fluid
kernel $c_{0}(r_{12};\rho _{0}^{\mathbf{r}_{1}})$ that depends implicitly on 
$\mathbf{r}_{1}$ through $\rho _{0}^{\mathbf{r}_{1}}$. This $\mathbf{r}_{1}$
dependence is the most important new feature of the HLR equation and it
represents the main reason for improved results over conventional methods,
which generally use only $c_{0}(r_{12};\rho _{0}^{B})$ along with various
nonlinear closures that try to directly relate the field and the density in
regions where the field is nonzero (10). See, e.g., Equation \ref{solutedcf}
below.

\subsubsection{Hard sphere solute and the PY equation}

Consider in particular the important test case where $\phi _{R}$ represents
the potential of a hard sphere solute fixed at the origin. The local
hydrostatic response $\rho _{0}^{\mathbf{r}_{1}}$ is simply a step function
in this case, equal to zero for $\mathbf{r}_{1}$ inside the core region and
to $\rho _{0}^{B}$ outside. In reality at high density there are very large
nonlocal oscillatory excluded volume correlations in the true $\rho _{0}(%
\mathbf{r}_{1})$. As we will see, the HLR equation gives results in this
special case equivalent to the generally accurate Percus-Yevick (PY)
approximation. In addition, we can use results from recent computer
simulations (75) and related work on the Gaussian field model (76) to
provide us with further insights into why this simple linear response
treatment can remain surprisingly accurate even for hard cores.

We can connect the HLR equation with conventional integral equation theory
leading to the PY equation by noting that the usual solute-solvent direct
correlation function $C_{S}(\mathbf{r}_{1})$ can be defined by 
\begin{equation}
C_{S}(\mathbf{r}_{1})=\int dr_{2}\chi _{0}^{-1}(r_{12};\rho _{0}^{B})[\rho
_{0}(\mathbf{r}_{2})-\rho _{0}^{B}],  \label{solutedcf}
\end{equation}
where exact values for all functions are used on the right side. Thus $%
-C_{S}(\mathbf{r}_{1})/\beta $ is that function that must replace the
perturbing solute-solvent potential to give exact results when the full
induced density change from $\rho _{0}^{B}$ is used in the linear response
equation \ref{linresponse}. (Using Equation \ref{chiinverse}, we see that
Equation \ref{solutedcf} is equivalent to the standard solute-solvent
Ornstein-Zernike equation, which is the usual way (10) of defining $C_{S}(%
\mathbf{\ r}_{1}).$)

Similarly, in the derivation of the HLR equation we can replace the zero on
the left side of Equation \ref{chilinhydro} by a new function $\widetilde{C}%
_{S}(\mathbf{r}_{1})$ that exactly satisfies: 
\begin{equation}
\widetilde{C}_{S}(\mathbf{r}_{1})=\int \!d\mathbf{r}_{2}\,\chi
_{0}^{-1}(r_{12};\rho _{0}^{\mathbf{r}_{1}})[\rho _{0}(\mathbf{r}_{2})-\rho
_{0}^{\mathbf{r}_{1}}].  \label{ctilde}
\end{equation}
Note that the local value of the perturbing field $\phi _{R}^{\mathbf{r}
_{1}} $ associated with $\widetilde{C}_{S}(\mathbf{r}_{1})$ always vanishes
at $\mathbf{r}_{1},$ unlike the case for $C_{S}(\mathbf{r}_{1})$ above. The
HLR equation \ref{HLR} follows from Equation \ref{ctilde} by setting $\rho
_{0}^{\mathbf{r}_{1}}\widetilde{C}_{S}(\mathbf{r}_{1})$ equal to zero
everywhere, and to the extent that the HLR equation is accurate, we expect $%
\rho _{0}^{\mathbf{r}_{1}}\widetilde{C}_{S}(\mathbf{r}_{1})$ to be generally
very small. In particular the HLR equation \emph{automatically} satisfies
the \emph{core condition} $\rho _{0}=0$ inside the hard core region where
both $\rho _{0}^{\mathbf{r}_{1}}$ and $\rho _{0}(\mathbf{r}_{1})$ vanish.
Corrections to the HLR equation arise from nonzero values of $\rho _{0}^{%
\mathbf{r}_{1}}\widetilde{C}_{S}(\mathbf{r}_{1})$ outside the core.

If we compare Equations \ref{solutedcf}, with the core condition imposed,
and \ref{ctilde} for $\mathbf{r}_{1}$ \emph{outside} the core region where $%
\rho _{0}^{\mathbf{r}_{1}}=\rho _{0}^{B},$ we see they are identical. Thus $%
\widetilde{C}_{S}(\mathbf{r}_{1})$ outside the core exactly equals $C_{S}(%
\mathbf{r}_{1}),$ the ``tail'' of the hard sphere solute-solvent direct
correlation. This is generally believed to be small and in the very accurate
GMSA equation it is approximated by a rapidly decaying Yukawa function (73,
74). In the HLR approximation this tail is set equal to zero. Thus the HLR
equation predicts that the hard sphere solute-solvent direct correlation
vanishes identically outside the core. As is well known, this is equivalent
to the PY closure for hard core systems (10). A GMSA treatment could be used
in Equation \ref{ctilde} to correct the HLR results if more accuracy is
needed. A self-consistent application of these ideas when the solute is the
same size as the solvent, requiring that $C_{S}=c_{0}=0$ outside the core,
yields the standard PY equation for hard spheres as interpreted by Stell
(77), with $c_{0}(r)$ equal to zero outside the core and
$\rho_{0}(r)=\rho _{0}^{B}g_{0}(r)$ equal to zero inside.
Similarly, the HLR equation reduces to the PY wall-particle equation when the
radius of the solute tends to infinity and the PY $c_{0}$ is used.

\subsubsection{Computer simulations and Gaussian fluctuations}

Further insight into the surprising accuracy of linear response theory for
uniform hard sphere fluids comes from recent computer simulations by Crooks
and Chandler (75), following related work on water (17). They have shown
that even \emph{large} spontaneous density fluctuations in a uniform hard
sphere fluid can be accurately described using the same Gaussian probability
distribution that describes \emph{small} fluctuations. Small density changes
induced by a small perturbing field are described by the linear response
equation \ref{linresponse}. By the fluctuation-dissipation theorem, the same
linear response function $\chi _{0}^{-1}(r_{12};\rho )$ controls the
spontaneous Gaussian density fluctuations in the uniform fluid (10, 76). In
particular, they considered the probability distribution for finding $N$
hard spheres in a spherical volume of the fluid and found that even \emph{%
cavity or void formation} with $N=0$ was
reasonably well described by the Gaussian
theory. Since a cavity influences the rest of the fluid in exactly the same
way as a hard sphere solute of the same size, the density response of the
hard sphere fluid to such a solute (or imposed region of zero density)
should indeed be well described using linear response theory with the
uniform fluid response function, as in Equation \ref{chilinhydro}. Thus a
key feature exploited in Equation \ref{HLR} is that density fluctuations in
the reference fluid are to a remarkable extent Gaussian. The simulation
results pertain to fluctuations in a uniform fluid. Similarly, to describe
effects of a general external field, the HLR equation considers
perturbations at each $\mathbf{r}_{1}$ to the uniform hydrostatic fluid
density $\rho _{0}^{\mathbf{r}_{1}}$ induced by the \emph{shifted} field $%
\phi _{R}^{\mathbf{r}_{1}}$, which vanishes at least locally at $\mathbf{r}
_{1}$.

\subsubsection{Other rapidly varying model potentials}

Thus we see that the HLR equation \ref{HLR} satisfies the following limits.
i) It is exact when $\phi _{R}(\mathbf{r})$ is very slowly varying and
exactly describes to linear order small corrections to the hydrostatic
approximation ii) It is exact for \emph{any} $\phi (\mathbf{r})$ at low
enough density, where there is a local relation between the potential and
induced density. iii) For a field $\phi (\mathbf{r})$ from a general hard
core solute, Equation \ref{HLR} reduces to the PY approximation, as
discussed above. To give more indications of the accuracy of Equation \ref
{HLR}, we review solutions (5) for some model potentials designed to show both
the strengths and the weaknesses of the present methods and compare with
computer simulations.

Figure~1 shows the correlation functions $g_{0}(r)\equiv \rho _{0}(r)/\rho
_{0}^{B}$ for a hard sphere system at a moderately high bulk density $\rho
_{0}^{B}=0.49$ induced by the deep attractive spherical parabolic
model potentials shown in the inset. (Reduced units, with distances measured
in units of the hard sphere diameter are used.) The HLR equation reproduces
the increased density inside the well, and the nonlocal oscillatory excluded
volume correlations, which show a local density \emph{minimum} at the center
of the well due to packing effects.

Since the external field enters Equation \ref{HLR} only locally through its
effect on $\rho _{0}^{\mathbf{r}_{1}},$ it is also easy to use it for the 
\emph{inverse problem} of determining the field associated with a given
density profile. As an example, the crosses in the inset gives the
potentials predicted by Equation \ref{HLR} given the simulation data for $%
g(r)$.

While these results are qualitatively very satisfactory for the most part, there
are some problems. Note in particular the slightly \emph{negative} density
at the center of the well; a positive density is not guaranteed in this
linear theory. (The related HM equation derived in (5) always yields a
non negative density and does slightly better here
but it performs less well in the hard wall limit.) One would
expect methods based on expanding about the uniform hydrostatic fluid to be
least accurate for potentials with very steep gradients, and poor results at
high density were found for a repulsive planar triangular barrier potential
that rose from $0$ to $10k_{B}T$ over a distance of one hard sphere
diameter. Results actually \emph{improve} as the barrier height increases
and the potential approaches a hard wall potential: the theory does better
for hard cores because there is no region of space where there is a large
gradient in the external field while at the same time the local density is
nonzero. A GMSA like treatment based on Equation \ref{ctilde} may improve
matters here. But large repulsive potentials with very steep gradients are
better treated by ``blip function'' methods (66) or other expansions about a
hard core system.

\section{Correcting the molecular field approximation}

\label{improvevdw} We now return to fluids with attractive interactions
as described using a self-consistently determined ERF.
Thinking about the hydrostatic limit also can help us
correct some quantitatively inadequate features of the simple MF
approximation \ref{mfeqn} for the ERF (7). In the limit of a uniform system
this describes all effects of attractive interactions in terms of the
constant VDW parameter $a.$ While this very simple approximation captures
much essential physics and gives a qualitative description of the uniform
fluid thermodynamic properties it certainly is not quantitatively accurate.
For example, when used to describe a slowly varying liquid-vapor interface,
it will predict shifted (molecular field) values for the densities of the
coexisting bulk liquid and vapor phases. The main problem with the theory is
not so much its description of the local density gradients, but its
predictions for the thermodynamic properties of the coexisting \emph{bulk}
phases (60).

To achieve quantitative agreement with bulk thermodynamic properties one can
replace the constant $a$ by a \emph{function} $\alpha $ that depends
(hopefully weakly, to the extent the classical VDW theory is reasonably
accurate) on temperature and density. This has been suggested many times in
the literature, usually in the context of perturbation theories for uniform
fluids (10), where approximate expressions giving such functions have been
derived (78). We review here a simpler and more empirical approach (7) that can
be used to determine $\alpha $ if an accurate analytic equation of state for
the bulk fluid is known, and we will use this in a simple modification of
the MF expression for the ERF to insure that \emph{exact} thermodynamic
results (consistent with the given uniform fluid equation of state) are
found in the hydrostatic limit of a very slowly varying ERF.

To that end, let us consider again the MF equation in the limit of a
constant field. Instead of using the MF \emph{approximation} for $\mu (\rho
) $ as in Equation \ref{mumf}, we assume that $\mu (\rho )$ is known from an
accurate bulk equation of state. In particular, we determine $\mu (\rho )$
from the 33-parameter equation of state for the LJ fluid (79) given by
Johnson, et al. This provides a very good global description of the stable
liquid and vapor phases in the LJ fluid and provides a smooth interpolation
in between by using analytic fitting functions. Thus it naturally produces a
modified ``van der Waals loop'' in the two phase region and seems quite
appropriate for our use here in improving the simplest MF description of the
uniform fluid.

Using known properties of the hard sphere fluid, we also have an essentially
exact expression for $\mu _{0}(\rho ).$ Using this we can relate
$\mu (\rho )$ and $\mu _{0}(\rho )$ in analogy to Equation \ref{mumf}: 
\begin{equation}
\mu (\rho )=\mu _{0}(\rho )-2\alpha (\rho )\rho  \label{mualpha}
\end{equation}
but with the constant $a$ replaced by a function $\alpha (\rho )$ of density
and temperature chosen so that Equation \ref{mualpha} holds. Since even the
simplest molecular field approximation is qualitatively accurate we expect
that the ratio $\alpha (\rho )/a$ will be of order unity and rather weakly
dependent on density and temperature.

We indeed found that the constant $a$ was a good overall compromise (7).
However the true $\alpha (\rho )$ fit to the equation of state exhibits
variations of up to about fifteen per cent as a function of
density and temperature, illustrating the need for an accurate equation
of state for quantitative accuracy.

Because of the strictly local response,
these results for a constant field can be used to determine exact
results in the hydrostatic limit of very slowly varying fields. We want to
modify Equation \ref{mfeqn} so that in the hydrostatic limit it will
reproduce these exact values, while still giving reasonable results for more
rapidly varying fields.

There is no unique way to do this, but the following simple prescription
seems very natural, and gives our final result, which we will call the
\emph{modified molecular field} (MMF) approximation for the ERF: 
\begin{equation}
\phi _{R}(\mathbf{r}_{1})-\phi (\mathbf{r}_{1})=\frac{\alpha (\rho _{0}^{%
\mathbf{r}_{1}})}{a}\int d\mathbf{r}_{2}\,\rho _{0}(\mathbf{r}_{2};\mathbf{[}%
\phi _{R}],\mu _{0}^{B})\,u_{1}(r_{12})+2\alpha (\rho ^{B})\rho ^{B}.
\label{mmfint}
\end{equation}
Thus the molecular field integral in Equation \ref{mfeqn} is multiplied by a
factor $\alpha (\rho _{0}^{\mathbf{r}_{1}})/a$ of order unity that depends
on $\mathbf{r}_{1}$ through the dependence of the hydrostatic density $\rho
_{0}^{\mathbf{r}_{1}}$ on the local value of the field $\phi _{R}(\mathbf{r}
_{1}),$ and the constant of integration $2a\rho ^{B}$ is replaced by the
appropriate limiting value of the modified integral. Note that the
hydrostatic density $\rho _{0}^{\mathbf{r}_{1}}$ remains smooth and
relatively slowly varying outside the core even when $\phi _{R}(\mathbf{r}%
_{1})$ contains a hard core. The nonlocal oscillatory excluded volume
correlations that can exist in the full density $\rho _{0}(\mathbf{r}_{1})$
do not appear in $\rho _{0}^{\mathbf{r}_{1}}$ because of the strictly local
response to $\phi _{R}$. Results using Equation \ref{mmfint} will be
discussed below.

\section{MVDW theory of nonuniform fluids with attractive interactions}

\subsection{Two step method}

\label{twostep}The ERF must be calculated self-consistently for fluids with
attractive interactions. Our new MVDW theory corrects the classical VDW theory
by using Equation \ref{HLR} to accurately determine $\rho _{0}(\mathbf{r}_{2})$
in Equation \ref{hydroerf}. Thus the MVDW theory requires the simultaneous
solution of \emph{two} equations: Equation \ref{hydroerf} (or the closely
related equation that arises if the Equation \ref{mmfint} is used for the
ERF), \emph{and} Equation \ref{HLR}. Equation \ref{hydroerf} determines
$\rho _{0}^{\mathbf{r}_{1}}$, the local hydrostatic response to the ERF
as in the VDW theory described above, and Equation \ref{HLR}
determines the full nonlocal response $\rho _{0}(\mathbf{r}_{1})$.
(Similarly, the VDW theory can be viewed as replacing Equation \ref{HLR} by
the hydrostatic approximation
$\rho _{0}(\mathbf{r}_{1})=\rho _{0}^{\mathbf{r}_{1}}$.)

One can think about solving these equations by
a \emph{two step iterative method} (3).
For a given approximation to the ERF, determine in a first step the
associated smooth hydrostatic density $\rho _{0}^{\mathbf{r}_{1}}$ from
Equation \ref{hydroerf}. Then in a second step take account of nonlocal and
usually oscillatory corrections to this profile, generally induced by
rapidly varying features in the external field $\phi ,$ using a locally
optimal application of linear response theory, Equation \ref{HLR}. This new
density $\rho _{0}(\mathbf{r}_{1})$ is then used to compute a new
approximation for the ERF, and the two steps are iterated to self
consistency. This process is easy to implement numerically and rapid
convergence has been found in most cases we have examined (3, 6, 7).

We noted in Section \ref{simplemf} that the simple MF ERF can quite
generally be written as $\phi _{R}=\phi +\phi _{s},$ where $\phi _{s}$ in
Equation \ref{phis} takes account of the unbalanced attractive interactions
and is smooth and relatively slowly varying. We are often interested in
cases where $\phi $ is a hard core potential. (If $\phi $ also has a slowly
varying part, say from weak attractive interactions between a solute and
solvent, the latter should be added (6) to $\phi _{s}$ in the discussion that
follows.) In this case one can implement the two step process in a slightly
different way, which is physically suggestive and was in fact how we first
thought about the problem (3, 4, 6).

Let us consider in the first step the hydrostatic response $\rho _{s}^{%
\mathbf{r}_{1}}$ to the slowly varying part $\phi _{s}$
of the ERF \emph{alone.} Since there is a strictly
local response to the ERF, $\rho _{s}^{%
\mathbf{r}_{1}}$ differs from $\rho _{0}^{\mathbf{r}_{1}}$ for the full ERF
only inside the core region, where $\rho _{0}^{\mathbf{r}_{1}}$ vanishes
while $\rho _{s}^{\mathbf{r}_{1}}$ remains continuous and smooth. This
smoothness allows us to use the gradient approximation in the next to last
term of Equation \ref{hydroerf1} in determining $\rho _{s}^{\mathbf{r}_{1}} $
if we wish. Since $\phi _{s}$ describes the unbalanced attractive
interactions associated with interface formation, we can interpret the
smooth $\rho _{s}^{\mathbf{r}_{1}}$ as an \emph{interfacial component} of
the full density response $\rho _{0}(\mathbf{r}_{1}).$

Then in the second step we take into account the response to the remaining
hard core part of $\phi _{R}$ and all nonlocal effects. This will cause
$\rho _{0}(\mathbf{r}_{1})\ $to vanish inside the core and, depending on the
extent to which the density $\rho _{s}^{\mathbf{r}_{1}}$ near the core has
been reduced, can induce nonlocal oscillatory
excluded volume corrections to $\rho _{s}^{\mathbf{r}_{1}}$
outside the core, which we again calculate using
linear response theory. Thus speaking pictorially, the second step takes
into account the nonlocal Gaussian fluctuations induced by the hard core
potential about the smoothly varying interfacial component of the density
profile. The strongly non-Gaussian component associated with interface
formation and arising from the unbalanced attractive interactions is taken into
account in the first step. The final profile results from the
self-consistent interplay between the components described in each step.

Since we are using linear response theory in all cases to correct a
hydrostatic response, these alternate ways of implementing the two step
procedure are completely equivalent. In particular in the first
interpretation described above we calculate the hydrostatic response $\rho
_{0}^{\mathbf{r}_{1}}$ to the full ERF and correct it for all $\mathbf{r}%
_{1} $ using Equation \ref{chilinhydro}. The hard core condition $\rho _{0}(%
\mathbf{r})=0$ for all $\mathbf{r}$ inside the core region follows \emph{%
automatically} from perturbing about the local hydrostatic density. In the
second interpretation we calculate in the first step the hydrostatic
response $\rho _{s}^{\mathbf{r}_{1}}$ to the interfacial component $\phi
_{s}$ alone and correct it by using linear response only for $\mathbf{r}%
_{1}$ \emph{outside} the core, while \emph{imposing}
$\rho _{0}(\mathbf{r}_{2})=0$ in the core region:
\begin{equation}
0=\int \!d\mathbf{r}_{2}\,\chi _{0}^{-1}(r_{12};\rho _{s}^{\mathbf{r}%
_{1}})[\rho _{0}(\mathbf{r}_{2})-\rho _{s}^{\mathbf{r}_{1}}].
\label{chilinhydrorhos}
\end{equation}
Since $\rho _{s}^{\mathbf{r}_{1}}$ equals $\rho _{0}^{\mathbf{r}_{1}}$
outside the core, the solution $\rho _{0}(\mathbf{r})$ to Equation \ref
{chilinhydrorhos} with the core condition imposed is identical to that
given by Equation \ref{chilinhydro}.

\subsection{Hard sphere solute in LJ\ fluid near liquid-vapor coexistence}

These ideas have been applied to the nonuniform LJ fluid in a variety of
different situations, including fluids near a hard wall, fluids confined in
slits and tubes, and the liquid-vapor interface, with generally very good
results (1-3, 5-7, 80-82). We will review here results (7) for a system
studied by Huang and Chandler (HC) that combines many of these limits (83),
and is physically very relevant for our later discussion about hydrophobic
interactions.

HC carried out extensive computer simulations to determine properties of the
LJ liquid at a state very near
the triple point with $\rho ^{B}=0.70$ and $T=0.85$
as the radius of a hard sphere solute is varied, and compared the results to
the LCW theory reviewed below. (We use the standard LJ reduced units.) By
definition the solute centered at the origin interacts with the LJ particles
through the hard core potential: 
\begin{equation}
\phi (r;S)=\left\{ 
\begin{array}{ll}
\infty , & r\leq S, \\ 
0, & r>S.
\end{array}
\right.  \label{phirS}
\end{equation}

The MMF theory discussed above allows us to reduce this problem to that of
the reference fluid in the presence of the effective field $\phi _{R}(r;S)$
satisfying Equation \ref{mmfint}. We have calculated self-consistently the
ERF $\phi _{R}(r;S)$ and the associated density response $\rho _{0}(r;S)$ of
the reference fluid, solving Equations \ref{mmfint} and \ref{HLR} by
iteration. In Figure~2 we compare these results for the density profiles in
the presence of the hard sphere solutes with $S$ equal to $1.0$, $2.0$, $3.0$
and $4.0$ in reduced units with the simulation results of the same LJ system
by HC. There is very good agreement between theory and simulation.

Figure~3 shows the corresponding ERF's obtained in these calculations. For
small solutes with $S$ less than about $0.7,$ attractive interactions do not
give rise to any substantial modification of the bare external field, as can
be seen from the plot of $\phi _{R}(r;S)$ for $S=0.5$. (Clearly, for $S=0$
there are no solute induced interactions of any kind and the profile reduces
to the constant $\rho ^{B}$). However, the effects due to unbalanced
attractions become important even for $S=1.0$, which is about the same size
as the LJ core, and all larger sizes give rise to a very strong and
relatively soft repulsion in $\phi _{R}(r;S)$. The corresponding density
profiles show pronounced depletion near the surface of the solute,
characteristic of surface induced drying.

\subsection{Solvation free energy}

Another quantity of great interest is the free energy of the nonuniform
system. This is the main focus of attention in density functional theory,
briefly discussed in Section \ref{DFTsection} below. In contrast, the MVDW
approach focuses first on the liquid \emph{structure}. We believe this
permits physical insight to play a more direct role. However, since we can
determine the density response to an arbitrary external field, the free
energy can be easily calculated from a coupling parameter type integration
that connects some initial state (e.g., the bulk fluid) whose free energy is
known to the final state as the field is varied. In the present case there
is a very simple route to the free energy of the nonuniform LJ system that
uses structural features that we know from simulations are accurately
determined.

In particular let us consider the change in free energy of the LJ fluid as
the range of the external field representing the hard core solute is varied
from zero to its full extent $S.$ This construction is the basis of scaled
particle theory (84), and it is well known that the free energy change takes
a particularly simple form: 
\begin{equation}
\beta \Delta \Omega _{S}=4\pi S^{3}\int\limits_{0}^{1}\!d\lambda \ \lambda
^{2}\rho _{\lambda }(\lambda S^{+}),  \label{HSomega}
\end{equation}
which requires only the \emph{contact value} $\rho _{\lambda }(\lambda
S^{+}) $ of the density profile. This is very accurately given by the theory
described above. To use this ``virial route'', we can replace the $\lambda $%
-integration by a sum and calculate the density profile for several values
of $\lambda $ at the fixed bulk chemical potential $\mu ^{B}$.

Using Equation \ref{HSomega} we obtain the dependence of solvation free
energy on the size of the hard sphere solute. The free energy per unit
surface area of the solute $\Delta \Omega _{S}/4\pi S^{2}$ we obtain is
shown in Figure~4. For small solutes unbalanced attractive interactions do
not play an important role and the solvation free energy agrees well with a
pure hard sphere model, which completely neglects attractions by using the
bare hard core solute potential, as shown by the dotted line. At the solute
size of about $0.7$ the behavior changes drastically and the reduced free
energy rapidly crosses over to the practically constant plateau in agreement
with the simulation results. The small slope of the curves in Figure~4 for
large $S$ can be understood by separating the free energy into volume ($%
V_{S}=4\pi S^{3}/3$) and surface ($A_{S}=4\pi S^{2}$) contributions as
discussed by HC (83): 
\begin{equation}
\Delta \Omega _{S}\;\approx \;V_{S\,}p^{B}+A_{S}\gamma _{S}.
\end{equation}
The first term in this expression corresponds to the work required to remove
liquid particles from the volume occupied the solute, where $p^{B}$ is the
bulk liquid pressure, and is very small for the values of $S$ considered
here. The second term determines the cost of forming the liquid-solute
interface and is proportional to the interface tension $\gamma _{S}$, which
is essentially independent of the solute size for large solutes.

\section{Hydrophobic interactions in water}

We now discuss an important extension of these ideas to hydrophobic
interactions in water (4), as described by Lum, Chandler and Weeks (LCW). We
first consider water at ambient conditions as the radius of a hard sphere
solute at the origin is varied. Because this state is very close to the
liquid-vapor phase boundary, phenomena involving interfaces can be very
important. This system serves as a simple model of a
hydrophobic object in water --- the solute does not
participate in hydrogen bonding
and it creates an excluded volume region where the density of water
molecules vanishes. Weak van der Waals attractions between the solute and
solvent do not change the qualitative nature of the phenomena we will
discuss and can easily be taken into account (6, 85).

But how can one sensibly apply the theory to water?
The local structure of water
is certainly very different from that of the LJ fluid and anything relying
on the detailed properties of a hard sphere reference system cannot be
trusted. However the two systems do have certain essential features in
common that can be exploited in a properly generalized MVDW theory.

First, small fluctuations in liquid water are Gaussian, and computer
simulations had earlier shown that even relatively large fluctuations
leading to the formation of molecular sized cavities can be well described
using the same Gaussian distribution (17). Indeed Pratt and Chandler (13)
developed a quantitative theory for the solvation of small apolar molecules
using the experimental linear response function for water that takes into
account the structural and free energy changes induced by the excluded
volume of the solute. These ideas have been significantly extended and
placed on a firmer conceptual basis in recent work by Pratt, Hummer and
coworkers (17, 19, 20, 22, 30, 86). See also (76). This suggests that the
Gaussian/linear response ideas used in the second step of the two step
method in the MVDW theory, using response functions appropriate for water,
could be modified to apply to water.

Second, just as for the LJ fluid above,
the main non-Gaussian feature to be
expected in this application is associated with \emph{interface formation}.
Long ago, Stillinger (11) gave a qualitative and physically very suggestive
description of what would be expected as the radius of the hard sphere
solute is increased. While small solutes should not significantly disturb
the hydrogen bond network, which can simply go around the solute, in the
vicinity of a sufficiently large solute or wall the network must be
completely disrupted. In the latter case, Stillinger argued, the arrangement
of molecules near the solute and the interface free energy should resemble
that of the liquid-vapor interface, which optimally solves the similar
problem of going from a complete hydrogen-bond network in the liquid to no
hydrogen bonds in the dilute vapor.

Thus Stillinger envisioned a drying transition very like that studied in the
last Section for the LJ fluid as the solute size is increased. This
phenomena is very general. The differences in local structure should have
little effect on the generic and qualitative physics leading to interface
formation. Thus it seems plausible that the interfacial component determined
in the first step of the MVDW theory could be described at least
qualitatively by a MF treatment similar to that used for the LJ fluid
provided appropriate thermodynamic parameters for water are used.

We see that key features of both steps
of the MVDW theory have some analogues for
water. The hard sphere reference fluid picture for the LJ fluid was incisive
in developing the general ideas leading to the MVDW theory. Given that
understanding, we may be able to develop an analogous approach for water and
other fluids that does not rely on the details of the reference fluid, or
indeed explicitly introduce a reference fluid at all.

Thus, following LCW, let us examine the essential features of the MVDW
theory and see how they can be modified to apply to water. As in the last
part of Section \ref{classicalvdw}, we will try to describe everything
formally in terms of the properties of water itself, and not a reference
system, though molecular field ideas will be introduced to define what is
meant in two-phase regions, etc. A crucial part of the physics of interface
formation in the MVDW theory is the description of the unbalanced attractive
interactions in Equation \ref{phis}. In the LJ fluid this can be rewritten
and reinterpreted in terms of an averaged or \emph{coarse grained density} $%
\overline{\rho }(\mathbf{r}_{1})$ with a normalized ``weighting function''
proportional to $u_{1}$: 
\begin{equation}
-2a\overline{\rho }(\mathbf{r}_{1})\equiv \int d\mathbf{r}_{2}\,\rho (%
\mathbf{r}_{2})\,u_{1}(r_{12}).  \label{rhobar}
\end{equation}
LCW argue that the unbalanced attractive interactions in water can be
described by a similar coarse graining of the water density, with the coarse
graining carried out over the appropriate range $\lambda $ of the attractive
interactions in water.

Now consider the first step of the two step method, as described in the last
part of Section \ref{twostep} where the smooth interfacial component $\rho
_{s}^{\mathbf{r}_{1}}$ associated with the field $\phi _{s}$ in Equation \ref
{phis} is determined from Equation \ref{hydroerf1}. Using the notation of
Equation \ref{rhobar} and expanding the next to last term, Equation \ref
{hydroerf1} becomes: 
\begin{equation}
\mu (\rho _{s}^{\mathbf{r}_{1}})=\mu ^{B}+m\nabla ^{2}\rho _{s}^{\mathbf{r}
_{1}}+2a[\overline{\rho }(\mathbf{r}_{1})-\overline{\rho }_{s}^{\mathbf{r}%
_{1}}].  \label{lcw5}
\end{equation}
With appropriate change of notation this is exactly Equation 5 of LCW.
However LCW did not consistently interpret $\rho _{s}^{\mathbf{r}_{1}}$ ($%
=n_{s}(\mathbf{r}_{1})$ in LCW) as the hydrostatic density and some later
simplifications that arise from this were not exploited. Since the slowly
varying and generic interfacial component $\rho _{s}^{\mathbf{r}_{1}}$ in
Equation \ref{lcw5} should be essentially independent of local structure,
LCW used a simple VDW form for $\mu (\rho ),$ which automatically
interpolates in the two phase region, but with VDW parameters $a$ and $b$
chosen to reproduce the density and compressibility of liquid water at phase
coexistence and $T=298$K. The parameter $m$ was fit to the surface tension
of water and the VDW relationship between $a$ and $m$ determined the
coarse-graining scale $\lambda $ for $\overline{\rho }(\mathbf{r),}$ which
LCW carried out using a simple Gaussian weight. While the details of this
fitting procedure are rather arbitrary, and the VDW equation is probably
inadequate to describe the quantitative relation between the energy density
and surface tension in water (63), LCW showed that small variations in these
parameter values did not change the qualitative picture that emerged.

Now turn to the second linear response step, described for the LJ\ fluid by
Equation \ref{chilinhydrorhos}. What is the analogue of the LJ reference $%
\chi _{0}^{-1}(r_{12};\rho )$ for water? In view of Equation \ref{pairden},
one can formally consider fluctuations in the full fluid instead. For $\rho $
in the stable liquid phase, small fluctuations are Gaussian and the \emph{%
uniform fluid} $\chi ^{-1}(r_{12};\rho )$ can be used directly. However
when interfaces form, we need an approximation for $\chi ^{-1}$ in our MF
treatment that remains well defined for all values of $\rho $ in the
two-phase region as well as in the vapor phase, as is the case for the
reference $\chi _{0}^{-1}$ for the LJ fluid. In effect LCW devised an
interpolation scheme for a $\chi ^{-1}$ for water that has these properties.

LCW essentially considered an alternate but equivalent version of Equation
\ref{chilinhydrorhos}: 
\begin{equation}
\rho (\mathbf{r}_{1})-\rho _{s}^{\mathbf{r}_{1}}=\int \!d\mathbf{r}%
_{2}\,\chi (r_{12};\rho _{s}^{\mathbf{r}_{1}})C_{S}(\mathbf{r}_{2}),
\label{chiwater}
\end{equation}
which involves the standard linear response function $\chi (r_{12};\rho
)\equiv \rho \delta (\mathbf{r}_{1}-\mathbf{r}_{2})+\rho ^{2}h(r_{12};\rho )$,
the inverse to $\chi ^{-1}(r_{12};\rho ).$ Here $h+1$ is the radial
distribution function for water and $C_{S}$ is a generalized solute-solvent
direct correlation function, similar to that in Equation \ref{solutedcf}.
This is nonzero only inside the core (PY closure), and is completely
determined by the imposed
requirement that $\rho (\mathbf{r})$ equals zero for all $\mathbf{r}$ in the
core region. They replaced $h(r_{12};\rho )$ by $h(r_{12};\rho ^{B})$ in the
definition of $\chi $, and used experimental values for the latter. This
approximation is exact in the bulk liquid phase, reduces to the correct
ideal gas value at very low density, and smoothly interpolates in between.
Equation 6 in LCW reduces to our Equation \ref{chiwater} when it is realized
that $\rho _{s}^{\mathbf{r}_{1}}$ is the hydrostatic density. Since $\rho
_{s}^{\mathbf{r}_{1}}$ is slowly varying, no essential differences should
result from using either equation.

Thus LCW implement in an approximate, but plausible way both steps of the
MVDW theory, making use of experimental thermodynamic and structural data.
They obtained results for the solute-solvent density distribution function
qualitatively very similar to that of Figure 2 for the analogous LJ-hard
sphere system. As the radius of the solute increased they found a crossover
on the biologically relevant length scale of \emph{nanometers }from
``wetting'' with peak densities greater than the bulk to ``drying'' with
peak densities less than the bulk. (The unbalanced attractive interactions
cause significant density perturbations in both regimes, so this terminology
is somewhat misleading.) Weak attractive van der Waals attractions between
the solute and water can \emph{shift} inward the position of the interface
when partial drying occurs and suppress complete drying, but should not have
an important effect on the basic interface structure and free energy changes
or the length scale for the crossover (85).

LCW also calculated solvation free energies, using a Gaussian approximation
that is quantitatively somewhat less accurate than the coupling parameter
method discussed above for the LJ system, but quite sufficient for
qualitative purposes. Again they found behavior very similar to Figure 4,
with free energies scaling with surface area only on large length scales of
order nanometers. LCW also studied assemblies of extended idealized
hydrophobic objects (plates and rods) and found that the drying can lead to
strong attractions between sufficiently close pairs of such surfaces as the
intervening water is expelled. Thus there is a \emph{length scale dependence}
of hydrophobic interactions. LCW suggest that such phase transitions
could play an important role in aspects of protein folding where
extended mostly hydrophobic regions approach one another.

It is beyond the scope of this article (and the expertise of this author!)
to assess the validity of that last conjecture, given the many complications
occurring in nature. Clearly much more theoretical and experimental effort
is called for on all aspects of the theory. Some interesting recent work
along these lines can be found in (21-29, 87-93). Our purpose is merely to
argue that interface formation is a fundamental piece of physics that
almost certainly occurs for the idealized models discussed herein and that LCW
have developed a qualitatively reasonable MF treatment of that process
based on sound statistical mechanical principles. From that perspective, the
scaling of hydrophobic solvation energies with exposed surface area with a
value close to the surface tension of water can be justified only on large
length scales (4, 93).

\section{Density functional theory}

\label{DFTsection}As discussed in the Introduction, the most commonly used
theory for the structure and thermodynamics of nonuniform fluids is \emph{%
density functional theory }(DFT). Space limitations permit only a few
general remarks here focusing on possible relations to the MVDW theory and
the advantages and disadvantages of each approach. Two standard reviews of
DFT are found in (43, 44).

A main focus of DFT is the \emph{intrinsic free energy density functional}
$F([\rho ],[w])$, which arises from the usual grand canonical free energy
$\Omega $ by a Legendre transform (10), where the functional dependence on
the chemical potential and external field $\phi (\mathbf{r})$ is replaced by a
functional dependence on the (uniquely associated) singlet density
$\rho (\mathbf{r})$.
Our notation emphasizes that $F$, just like $\Omega ,$ remains a functional
of the intermolecular pair (and any higher order) potentials $w.$ In
principle this formalism is exact and $F$ contains all the information in
$\Omega .$ An exact hierarchy of (direct) correlation functions can be
derived by successive functional derivatives of $F$ with respect to $\rho $.

While it must be as hard to determine $F$ \emph{exactly} as it is $\Omega ,$
the hope is that practically useful approximations may suggest themselves
more naturally in this representation, and the sound theoretical basis of
DFT may provide a means for systematic corrections. By starting from the
free energy, certain exact \emph{sum rules} relating integrals of those
correlation functions to the thermodynamic properties are \emph{automatically}
and consistently satisfied in any DFT (45, 59, 60); this is not the case
for structurally based methods such as the MVDW theory. Particularly
suggestive is the fact that the classical VDW theory can be viewed as a DFT
for $F$ where a \emph{MF approximation} is made for the functional
dependence on $w$ and a \emph{local density approximation} is made for the
functional dependence on $\rho $. While this certainly seems promising, it
turns out to be rather difficult to do significantly better from this
starting point.

Thus, can the DFT formalism help us improve on the MF approximation?
Consider the first functional derivative of $F,$ which from basic properties
of the Legendre transform is easily seen to satisfy 
\begin{equation}
\delta F([\rho ],[w])/\delta \rho (\mathbf{r})=\mu -\phi (\mathbf{r}).
\label{dFdrho}
\end{equation}
In principle if we knew the ``exact'' $F$ we could solve this equation,
determining the density $\rho (\mathbf{r})$ associated with a
given potential
$\phi (\mathbf{r}).$ While most discussions have focused on the density
dependence, the main problem in determining an accurate $F$ from a
fundamental point of view is its dependence on $w.$ When there are
attractive interactions, the exact $F$ must describe critical phenomena,
capillary waves, and a host of other properties for which we have
essentially no idea what the true functional dependence on $w$ should be. A
treatment incorporating even the simplest capillary wave correlations for
the liquid-vapor interface produced a density functional very different from
conventional ideas (44, 72, 94), and this does not begin to address the
range of phenomena arising from the general functional dependence on $w.$ So
as a practical matter we are essentially forced to accept the MF treatment
of the attractive interactions, perhaps modified slightly as in Section \ref
{improvevdw} to give a better description of the uniform fluid (60).

The true $F$ for a system with repulsive forces only must be significantly
simpler. Even here the dependence on a general repulsive $w_{0}$ can cause
problems; any density functional must also explicitly or implicitly
approximate the dependence on $w_{0}$. It is not obvious from a fundamental
point of view how to modify a functional that gives a good description
for say hard spheres, so that it can describe much softer repulsions.
See (53, 54) for interesting work on this question.
Unlike most quantum-mechanical applications
of DFT, where there is a single coulomb interaction potential, we must deal
with a range of interactions and hence potentially many different functionals.

Most workers have quite reasonably focused on the basic hard sphere system,
and here some significant progress has been made in determining accurate
approximations for $F_{0}$, particularly in the
development of fundamental measure theory (50-52, 58). However, the latter
is very complicated and one might hope for a simpler and perhaps more
physically motivated approach.

Unfortunately, the DFT formalism itself gives few indications of how to
proceed in a practical manner. The basic problem can be seen when we
consider Equation \ref{dFdrho} for a system of hard spheres and a general
$\phi _{R},$ i.e., 
\begin{equation}
\delta F_{0}([\rho _{0}],[u_{0}])/\delta \rho _{0}(\mathbf{r})=\mu
_{0}^{B}-\phi _{R}(\mathbf{r}).  \label{dF0drho}
\end{equation}
This equation can be viewed as a generalization of our hydrostatic equation 
\ref{hydromudef}, which in principle could be used to
determine the exact density if a
proper $F_{0}$ could be found. Indeed Equation \ref{hydromudef} follows from
Equation \ref{dF0drho} for a very slowly varying profile when a
\emph{local density} approximation is made. But how should this be
corrected for rapidly varying densities? To go beyond the local density
approximation, various \emph{weighted density approximations} have been
proposed. However, from a fundamental point of view except in special
low dimensional cases (58) it is not even clear
that $F_{0}$ can be written as a simple functional of weighted densities;
certainly the original $\Omega _{0}$ is not a functional of a weighted $\phi
_{R}(\mathbf{r})!$ This contrasts with the development of the MVDW theory
from the VDW theory. There the simplest correction (linear response) is
clearly correct for small perturbing potentials, and it remains reasonably
accurate even in the hard core limit, reducing naturally to the PY
approximation. Usually in DFT this limit is imposed by hand.

Since many treatments of DFT have discussed the formal
advantages of the method, we have mostly focused here on
some of the difficulties as we see them. Despite
these (perhaps pedantic!) objections, there have been impressive successes
arising from DFT, and there clearly are close connections between some of
the ideas of DFT and the MVDW theory. In particular, while we do think about
molecular \emph{fields}, the HLR equation \ref{HLR} explicitly involves only
densities. Indeed the reason for its success is the removal of any explicit
dependence on the field through the expansion about the hydrostatic density.
However, a deeper connection to DFT has so far escaped us; this is an active
topic of our current research.

\section{Final Remarks}

We conclude with a few general remarks. The MVDW theory combines in a
self-consistent way two standard and widely used ingredients: molecular (or
mean) field theory and linear response (or Gaussian field) theory. The
molecular field equation for the ERF takes account of the unbalanced
attractive forces discussed in the original VDW theory (9); all that is
required is to determine the molecular field and the induced structure
accurately. While the hydrostatic approximation used in our interpretation of
the VDW theory is not generally accurate, it serves as an optimal starting
point for corrections based on linear response theory. The resulting theory
can handle problems involving
fluctuations on a variety of length scales where each ingredient alone would
fail. Because of its sound physical basis, we are hopeful that the MVDW
theory will prove useful in many different applications. In particular,
assessing and further developing both the physical and biophysical
implications of the LCW theory in realistic environments seems an important
topic for future research. One new direction we are thinking about involves
fluids with long-ranged coulomb forces. Here there are a wide range of
phenomena such as charge ordering and double layer formation arising from
very strong and competing interactions on many length scales. This will
certainly put our current ideas to a most severe test.

There are also many basic theoretical questions left open. Can the MVDW
theory be understood as some kind of weighted DFT? Is it possible to extend
these ideas to solid-fluid interfaces? Can one go significantly beyond the
molecular field picture in describing the effects of attractive interactions
while still maintaining a tractable theory? The LCW theory, despite its use
of properties only of water, is definitely a molecular field theory. But the
idea of short wavelength Gaussian fluctuations related to local structure
and long wavelength slowly varying fluctuations related to interface
formation seems more general. A reinterpretation of LCW theory from this
perspective is found in (63). We close with what is probably the most basic
question: Is it is generally correct to imagine that
fluctuations are essentially Gaussian
in most non-critical liquids except when interfaces form? A deeper
understanding of whether and why this is true is called for. If this
physically suggestive picture remains valid, then the MVDW theory has
captured in a surprisingly simple way much of the non-critical physics of
the liquid state.

\section*{Acknowledgments}

This work was supported by the National Science Foundation through Grants
CHE-9528915 and CHE-0111104. Significant parts of the work described here were carried
out by Jeremy Broughton, Robin Selinger, Katharina Vollmayr-Lee, and Ka Lum.
Kirill Katsov played the key role in the conceptual development of the HLR equation
and its numerical solution, and David Chandler's ideas and insights
shaped the LCW theory. We are grateful to Michael Fisher and Jim Henderson for
many helpful discussions.


\newpage

\begin{figure} [tbp]
\centerline{
\epsfxsize=6.0in
\epsfbox[60 30 680 530]{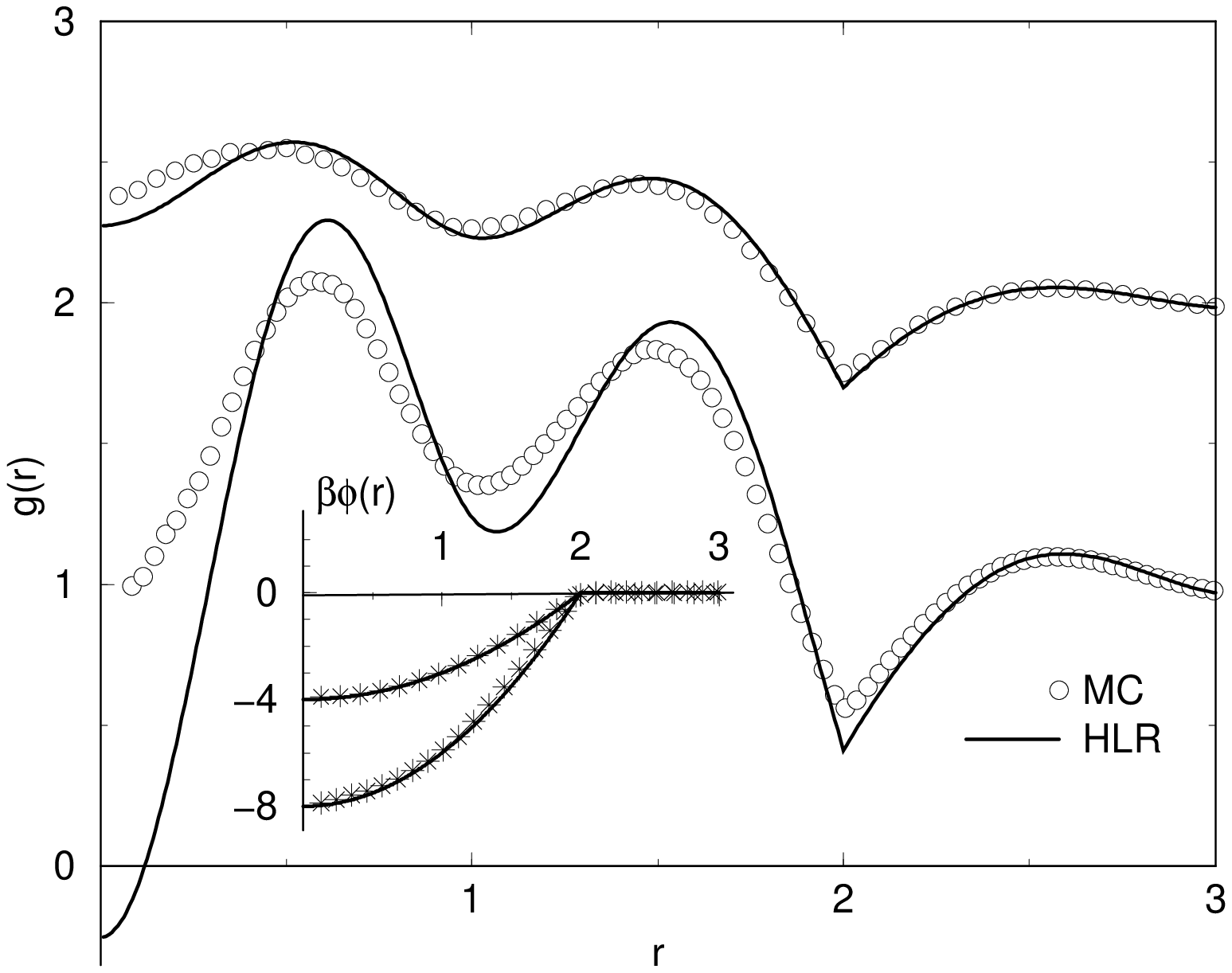}}
\caption{Correlation functions for hard spheres in the presence of
spherical parabolic potentials shown in the inset (solid lines) as given by
theory and simulation. The upper curve corresponds to the smaller potential
and has been displaced upward by one unit. Also shown in the inset (crosses)
are the potentials predicted by Equation \ref{HLR} given the simulation
data.}
\end{figure}

\begin{figure} [tbp]
\centerline{
\epsfxsize=6.0in
\vspace{0.6cm}
\epsfbox[60 30 680 530]{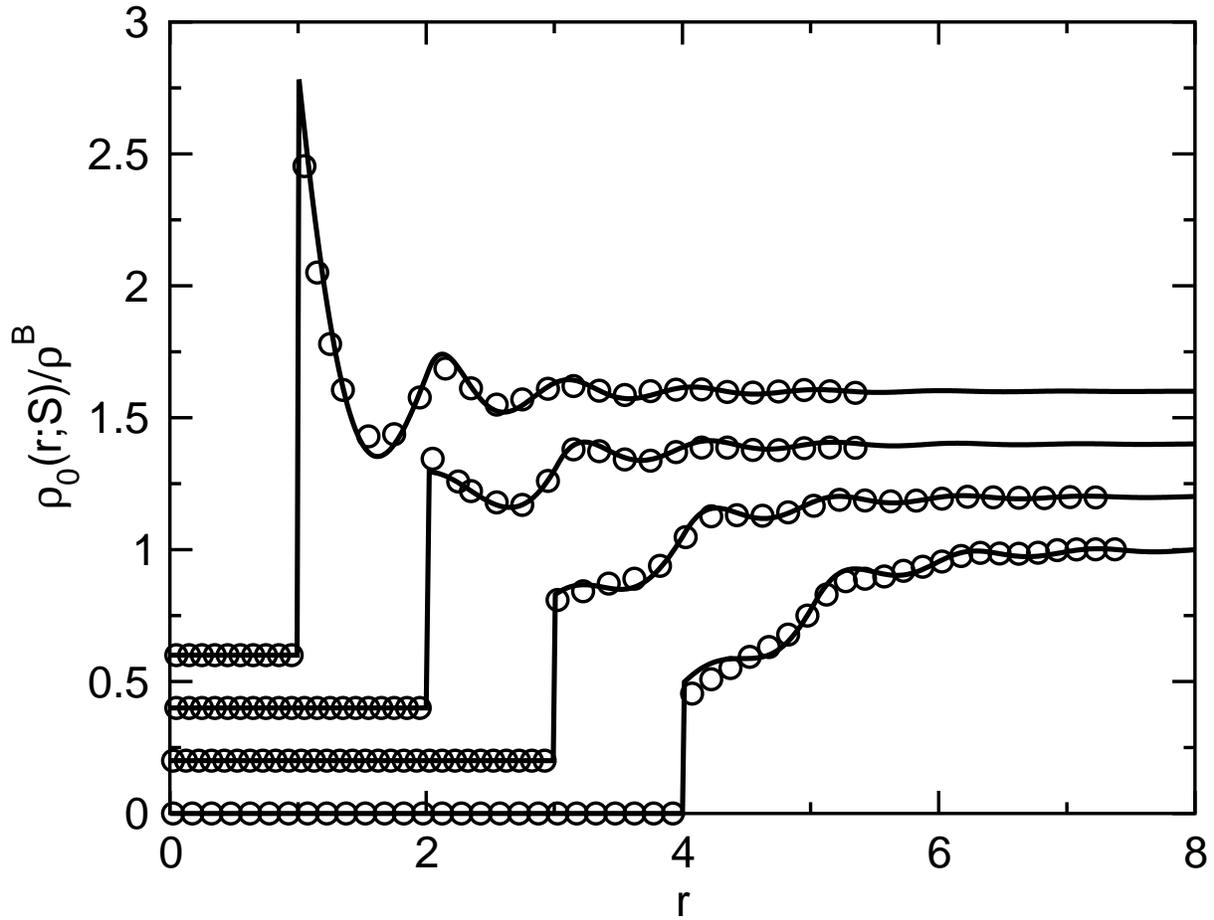}}
\caption{Density profiles of the LJ fluid ($T=0.85$, $\rho^{B}=0.70$)
in the presence of the hard sphere solute with $S=1.0$, $2.0$, $3.0$ and $4.0$.
Circles denote simulation results (83).
Lines are results of the self-consistent approach based on the modified
molecular field determined from Equation \ref{mmfint}.
For ease of viewing, the density profiles for $S=1.0$, $2.0$ and $3.0$
have been shifted vertically by $0.6$, $0.4$ and $0.2$ units respectively.}
\end{figure}

\begin{figure} [tbp]
\centerline{
\epsfxsize=6in
\vspace{0.2cm}
\epsfbox[60 40 680 530]{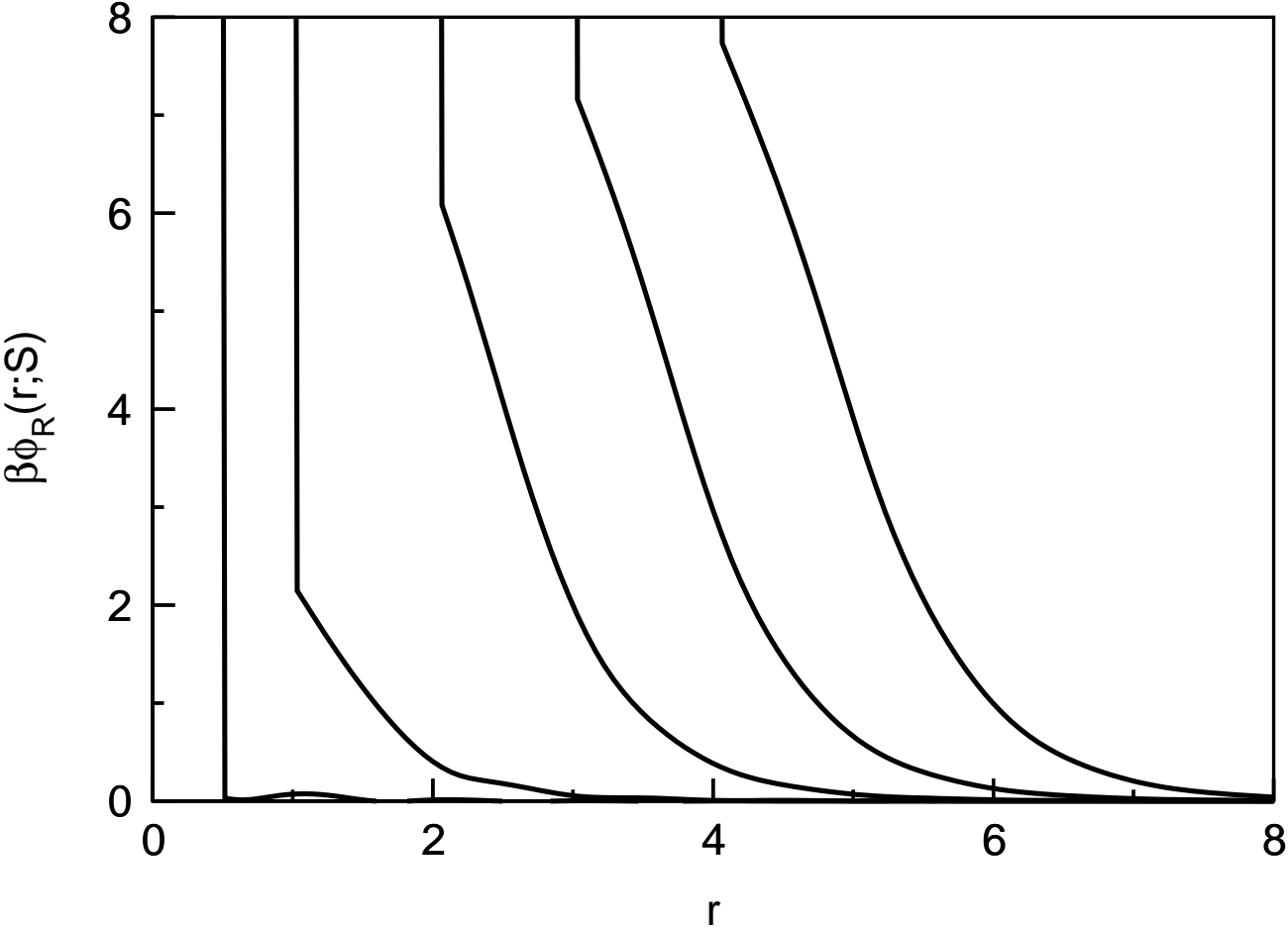}}
\caption{Self-consistent molecular field of the LJ fluid for the solute
with $S=0.5$, $1.0$, $2.0$, $3.0$ and $4.0$,
obtained from Equation \ref{mmfint}.}
\end{figure}

\begin{figure} [tbp]
\centerline{
\epsfxsize=6in
\vspace{0.6cm}
\epsfbox[60 60 680 530]{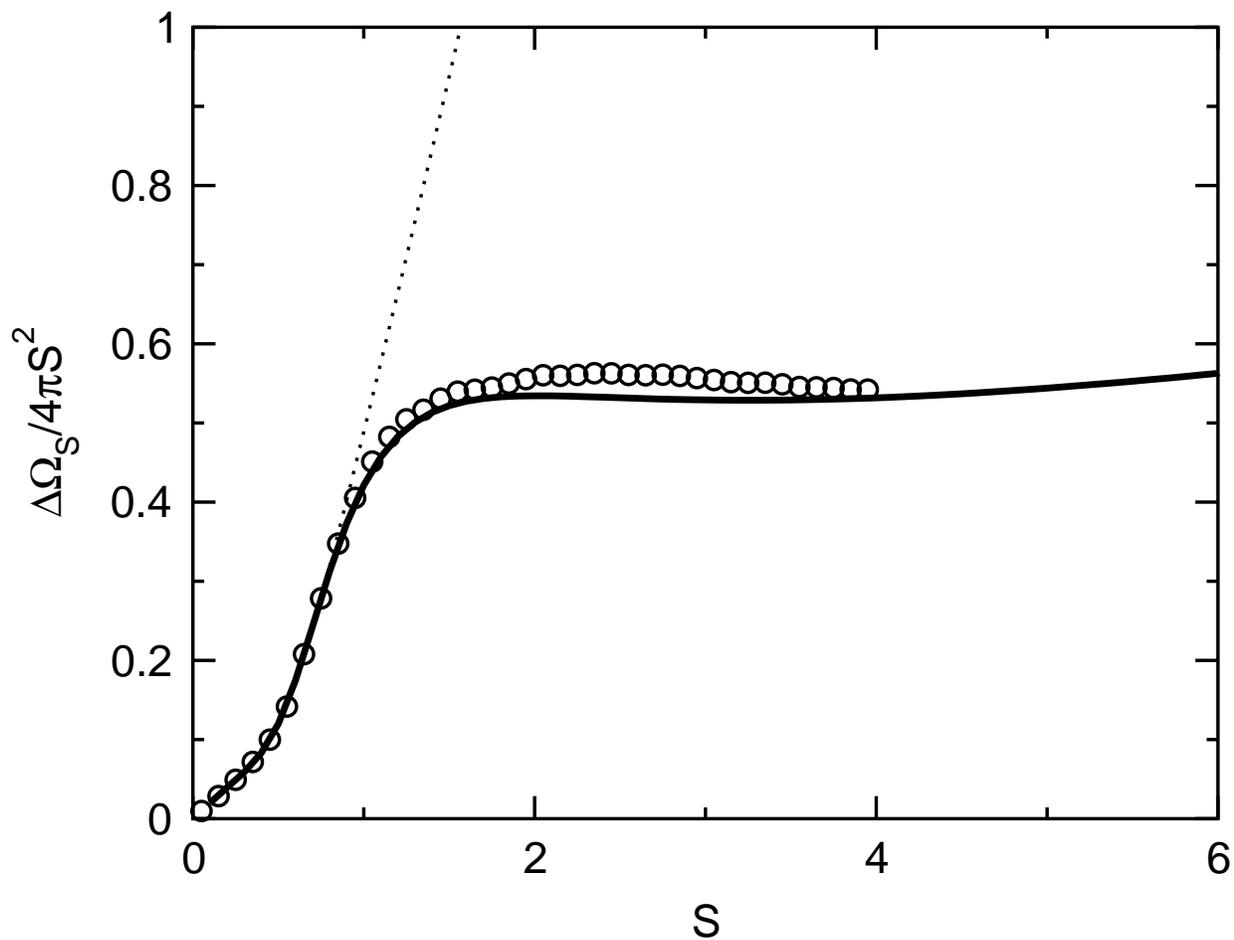}}
\caption{Dependence of the
solvation free energy on the cavity size $S$. Circles denote results of
simulations (83). Lines are obtained from Equation \ref{HSomega}
by using the results of the molecular field equation \ref{mmfint} (solid), and
by neglecting the molecular field (dotted).}
\end{figure}

\end{document}